\documentclass[reprint, amsmath, amssymb, aps, prb,superscriptaddress,longbibliography,floatfix]{revtex4-1}
\usepackage[english]{babel}
\usepackage[utf8]{inputenc}
\usepackage{amsmath}
\usepackage{mathtools}
\usepackage{lmodern}
\usepackage{lmodern}
\usepackage{bm}
\usepackage{dcolumn}
\usepackage{natbib}
\usepackage{graphicx}
\usepackage{soul}

\usepackage[breaklinks=true,colorlinks,citecolor=blue,linkcolor=blue,urlcolor=blue]{hyperref}
\usepackage{babel}
\def\be{\begin{equation}}
\def\ee{\end{equation}}
\def \bea{\begin{eqnarray}}
\def \eea{\end{eqnarray}}

\def \moire{moir\'e }
\usepackage{xcolor}

\begin{document}
\title{Nonlinear anomalous Hall effects probe topological phase-transitions in \\ twisted double bilayer graphene}
\author{Atasi Chakraborty}
\email{atasic@iitk.ac.in}
\affiliation{Department of Physics, Indian Institute of Technology Kanpur, Kanpur-208016, India}
\author{Kamal Das}
\email{kamaldas@iitk.ac.in}

\affiliation{Department of Physics, Indian Institute of Technology Kanpur, Kanpur-208016, India}

\author{Subhajit Sinha}
\affiliation{Department of Condensed Matter Physics and Materials Science, Tata Institute of Fundamental Research, Homi Bhabha Road, Mumbai 400005, India.}

\author{Pratap Chandra Adak}
\affiliation{Department of Condensed Matter Physics and Materials Science, Tata Institute of Fundamental Research, Homi Bhabha Road, Mumbai 400005, India.}

\author{Mandar M. Deshmukh}
\affiliation{Department of Condensed Matter Physics and Materials Science, Tata Institute of Fundamental Research, Homi Bhabha Road, Mumbai 400005, India.}

\author{Amit Agarwal}
\email{amitag@iitk.ac.in}
\thanks{\\A.C. and K.D contributed equally to this work.}
\affiliation{Department of Physics, Indian Institute of Technology Kanpur, Kanpur-208016, India}

\begin{abstract}

Nonlinear anomalous Hall effect is the Berry curvature dipole induced second-order Hall voltage or temperature difference in response to a longitudinal electric field or temperature gradient. These are the prominent Hall responses in time reversal symmetric systems. Here, we investigate the family of second-order nonlinear anomalous Hall effects, the electrical, thermoelectric and thermal nonlinear Hall effects in the \moire system of twisted  double bilayer graphene. We demonstrate that the nonlinear anomalous Hall signals can be used to probe the  topological phase-transitions in \moire systems, induced by a perpendicular electric field. Specifically, we show that the  whole  family of nonlinear anomalous Hall responses undergo a sign reversal across a topological phase-transition. 
\end{abstract}

\maketitle

\section{Introduction}


The experimental realization of novel quantum phases in small angle twisted bilayer graphene (TBG)~\cite{bistritzer_PNAS2011_moire,cao_N2018_uncon, cao_N2018_corre} has propelled \moire systems to the forefront of quantum material research. For very small twist angles, the enhanced impact of electron-electron interactions due to narrow bandwidth enables a rich phase diagram of correlated phases in TBG\cite{cao_N2018_corre,cao_N2018_uncon,serlin_Sc2020_intrin,sharpe_Sc2019_emergent}. In addition to TBG, other stacked \moire systems such as twisted double bilayer graphene (TDBG)~\cite{burg_PRL2019_corre,liu_N2020_tunable,cao_N2020_tunable, zhang_NC2021_visu, koshino_PRB2019_band, rodriguez_PRR2020_floquet,lee_NC2019_theory,he_NP2021_symmetry,kuiri_arxiv2022_spontaneous,sinha_NC2022_berry}, twisted trilayer graphene~\cite{phong_PRBL2019_band,zhu_PRL2020_twisted,zeyu_S2021_elec,park_N2021_tun}, and marginally twisted transition metal dichalcogenides~\cite{li_N2021_quantum,tran_N2019_eviden,weston_NN2020_atomic} also show fascinating strong correlation effects.
\begin{figure}
    \centering
    \includegraphics[width=.9\columnwidth]{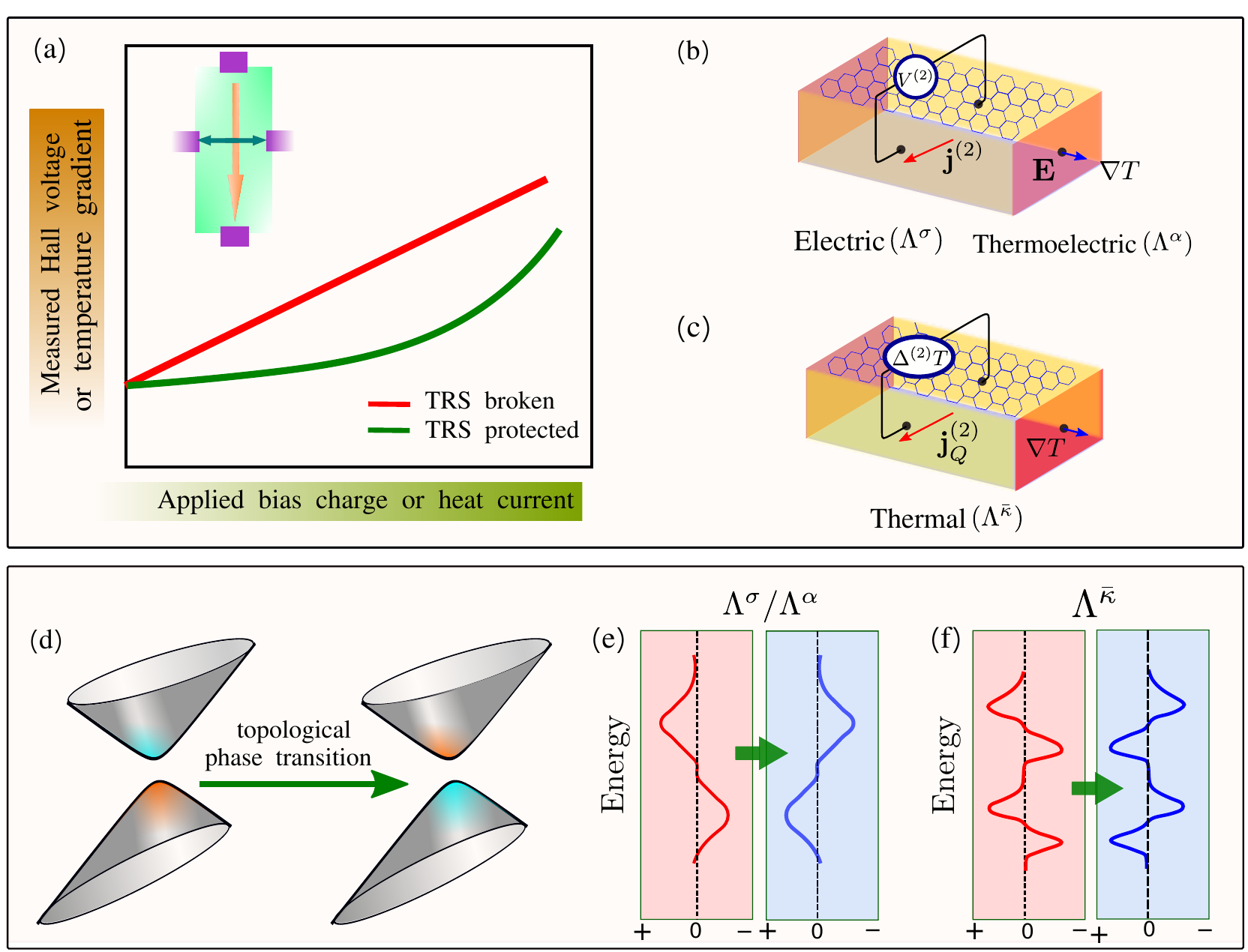}
    \caption{(a) The nonlinear (linear) anomalous Hall responses dominate the charge and heat Hall transport, shown in green (red) curve, in time reversal symmetry preserving (broken) systems. 
    (b), (c) We study the family of nonlinear anomalous Hall effect in strained twisted double bilayer graphene. These nonlinear Hall effects can be used to detect topological phase-transition. (d) The change of the Berry curvature distribution near the band edges during the topological  phase-transition. The impact of the topological phase transition on (e) the  nonlinear anomalous electrical and Nernst Hall effects and on (f) the thermal Hall effect. BCD ($\Lambda^\sigma$) and it's electrical ($\Lambda^\alpha$) and thermal ($\Lambda^{\bar \kappa}$) counterparts undergo a sign reversal across the phase-transition.} 
    \label{fig1}
\end{figure}
%
%
Moir\'e systems also offer a promising platform for exploring the valley and topological physics. In addition to large Berry curvature hotspots, valley-Chern number based topological phase transitions~\cite{zhang_PNAS2013_valley} have also been demonstrated in \moire systems such as TBG~\cite{zhang_PRB2019_nearly}, TDBG~\cite{koshino_PRB2019_band,chebrolu_PRB2019_flat,mohan_PRB2021_tri,wang_PRB2021_phase,sinha_NC2020_bulk,wang_NP2022_bulk,sinha_NC2022_berry}, multi-layered  twisted systems~\cite{liu_PRX2019_quantum,lin_PRB2021_emer}, and twisted transition metal dichalcogenides~\cite{hu_arxiv2020_non, huang_arxiv2021_giant}. 

The non-interacting topological aspects of \moire systems are relatively less explored. This is because the presence of time reversal symmetry (TRS) suppresses the Berry curvature related linear response phenomena, such as the anomalous Hall effect~\cite{nagaosa_RMP2010_anomal,sinitsyn_JPCM2007_semi}, in these materials. However, the nonlinear anomalous (NLA) Hall effect induced by the Berry curvature dipole (BCD)~\cite{sodemann_PRL2015_quantum,moore_PRL2010_confin,du_NRP2021_nonlinear,ortix_AQT2021_nonlinear,battilomo_PRL2019_berry,ho_NE2021_hall} can probe the Berry curvature in systems with TRS [Fig.~\ref{fig1}(a)]. The NLA Hall effect requires the breaking of space inversion symmetry (SIS), and it appears in crystals with reduced symmetry \cite{kang_NM2019_nonlinear,ma_N2019_obser,tiwari_NC2021_giant,you_PRBR_berry, zhang_2Dmat2018_electri,du_PRL2018_band,zhang_arxiv2020, sonia_PRBL2021_twist,pantaleon_PRB2021_tunable,hu_arxiv2020_non, huang_arxiv2021_giant,he_npjQM2021_giant}. 
In addition to the NLA Hall charge response to the electric field, there are also thermoelectric and thermal NLA responses that have a similar origin. The former is known as the NLA Nernst effect~\cite{yu_PRBR2019_topological,wu_PRB2021_nonlinear}, and the latter is termed as the NLA thermal Hall (or Righi-Leduc) effect~\cite{zeng_PRR2020_fundamental}. The family of the NLA charge and heat responses originate from the finite Berry curvature of the electronic bands, and the asymmetry of the electronic dispersion. 

In this paper, we demonstrate that the electrical, thermoelectric and thermal NLA Hall responses in strained TDBG can probe the topological phase transition, as they undergo a sign reversal across the transition. 
We present a systematic study of the family of NLA Hall effects: the electrical NLA Hall, the thermoelectric NLA Nernst~[Fig.~\ref{fig1}(b)] and the NLA thermal Hall~[Fig.~\ref{fig1}(c)] effect in strained TDBG. In particular, we highlight the impact of the band structure tunability of TDBG through the perpendicular electric field on the whole family of NLA Hall effects. While the electrical NLA Hall effect has been studied earlier in other moir\'e systems~\cite{zhang_arxiv2020, sonia_PRBL2021_twist,pantaleon_PRB2021_tunable, hu_arxiv2020_non, huang_arxiv2021_giant}, the thermoelectric and thermal NLA Hall effects have not been explored earlier in moir\'e systems.  
We connect the family of NLA Hall responses to the topological phase-transitions of the valley-Chern number~\cite{zhang_2Dmat2018_electri,facio_PRL2018_strong,hu_arxiv2020_non} in TDBG. 
In addition to tuning the electronic structure, the perpendicular electric field in TDBG also induces topological phase-transition of the valley-Chern type. We find that such transitions, accompanied by the band gap closing at a specific ${\bm k}$ point, change the distribution of the Berry curvature hotspot in the vicinity of the band crossing point~[Fig.~\ref{fig1}(d)]. The change in the Berry curvature distribution results in sign change of all three NLA Hall responses across the phase transition. This establishes the NLA Hall responses as a unique probe of the band topology in \moire and other topological systems.

\section{Electronic band structure of TDBG}

\label{band_TDBG}

TDBG is fabricated by placing two sheets of Bernal stacked (AB-stacked) bilayer graphene on top of each other and rotating them by an angle $\theta$ with respect to each other around an axis perpendicular to the plane. Accordingly, its Hamiltonian is constructed by combining the Hamiltonian of the individual bilayers along with a coupling between the two, known as moir\'e coupling. To obtain the electronic structure of TDBG, here we extend the low energy continuum approach of Bistritzer and MacDonald~\cite{bistritzer_PNAS2011_moire}. We particularly focus on the perpendicular electric field tunability and strain induced symmetry reductions of the TDBG which play pivotal roles in generating NLA Hall effects in TDBG. We emphasize that a perpendicular electric field can be experimentally applied in a specially designed field-effect-transistor like device structure with dual gates~\cite{he_NP2021_symmetry,kuiri_arxiv2022_spontaneous,sinha_NC2022_berry}. Strain, on the other hand, generally appears in 2D systems during fabrication on a substrate and is ubiquitous in almost every moir\'e  platform~\cite{kazmierczak_NM2021_strain}.

\begin{figure}[t]
    \centering
    \includegraphics[width=0.9\columnwidth]{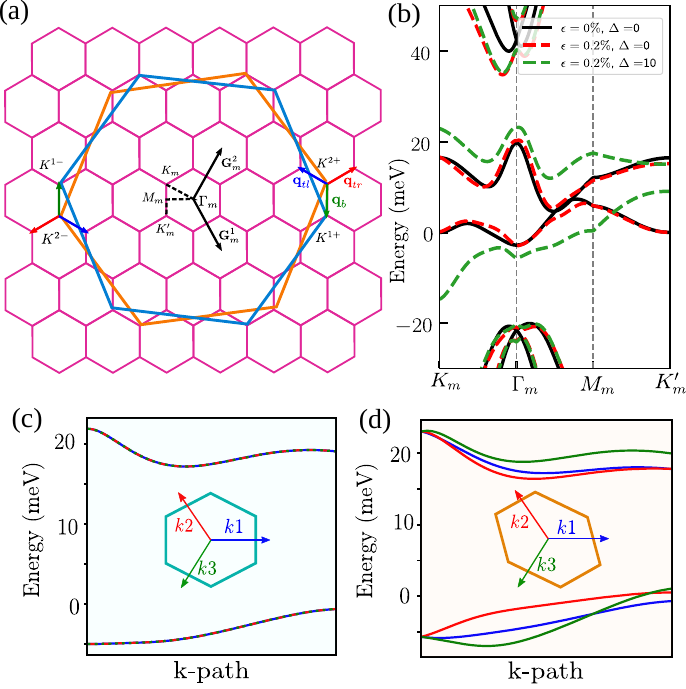}
    \caption{ (a) Small pink hexagons represent the moir\'{e} Brillouin zone (BZ). Large cyan (orange) hexagon is for first (second) bilayer graphene. Black arrows represents the moir\'e reciprocal lattice vectors and red, blue and green arrows represent nearest neighbors. The arrow orientation on the right side represents the $K$-valley (in our choice), and the left orientation represents the $K'$-valley. The dashed line represents the high-symmetry path in the \moire Brillouin zone.  (b) The energy dispersion of TDBG along the high symmetry points in presence and in absence of strain. Metal to insulator transition due the effect of electric field can be clearly seen. Dispersion of the pair of flat bands in the unstrained (c) and strained (d) cases along the three $C_3$ symmetric paths shown in the hexagon, highlighting the strain induced breaking of the $C_3$ symmetry in TDBG. 
    }
    \label{fig2}
\end{figure}

With the Bernal stacked bilayer graphene Hamiltonian, TDBG can be constructed to be either of AB-AB or AB-BA type~\cite{koshino_PRB2019_band,chebrolu_PRB2019_flat,mohan_PRB2021_tri,wang_PRB2021_phase}. Here, we consider the AB-AB configuration. However, we emphasize that the physics we highlight in this paper is independent of the stacking order. We assume that the bilayer at the top (layer-$1$) is rotated by an angle $-\theta/2$ and the bilayer at the bottom (layer-$2$) is rotated by an angle $\theta/2$. Such rotation of individual layers creates a moir\'e pattern with an effective twist angle $\theta$. A schematic of the moir\'e structure in the reciprocal space is presented in Fig.~\ref{fig2}(a). The big hexagons (orange and cyan) represent the rotated BZ of the (top and bottom) bilayer graphene and the small hexagons represent the moir\'e BZ. The neighbouring Dirac points (vertices of the large hexagons) of the rotated bilayers are connected by ${\bm q}_b, {\bm q}_{tr}$ and ${\bm q}_{tl}$. The details of the low energy model based band structure calculation of TDBG are summarized in Appendix~\ref{BS_wo_strain}.

The band structure of TDBG for twist angle $\theta=1.1^\circ$ is shown in Fig.~\ref{fig2}(b). In the absence of a perpendicular electric field, the band structure consists of a pair of low energy flat bands separated from the higher energy bands. The higher energy bands are referred to as the \moire bands and the corresponding gap is known as the \moire gap. The broad features of the electronic spectrum is similar to that found in TBG~\cite{bistritzer_PNAS2011_moire} and to other \moire systems~\cite{hu_arxiv2020_non}. However, TDBG offers better tunability of its electronic properties, which are accessed by the external perpendicular electric field [onsite parameter $\Delta$ in Eq.~\eqref{tdbg_ham}], compared to TBG.  We have included that in our calculations through the gap parameters as  $\Delta^{+}_t$=$-\Delta^{-}_b$=$\frac{3}{2} \Delta$ and $\Delta^{+}_b$=$-\Delta^{-}_t$=$\frac{1}{2}\Delta$ which basically represents a constant gradient in the potential. With this parameter $\Delta$, we can induce metal-insulator transition~\cite{adak_PRB2020_tunable} with out changing doping, along with topological phase-transitions of the valley-Chern type~\cite{zhang_PNAS2013_valley}. The topological phase transitions are associated with closure of the band gaps at a specific ${\bm k}$-point and it plays a very fundamental role in NLA Hall effects.

We now focus on the effect of strain on the band structure of TDBG.
Strain usually appears in 2D moir\'e systems when mounting on a substrate. In case of twisted \moire systems where two layers are put on top of each other and sandwiched between the substrates, strain may appear on both of the layers~\cite{bi_PRB2019_designing,pantaleon_PRB2021_tunable}. However, for simplicity, in our calculation we assume strain ($\mathcal E$) acts only on a single (bottom) layer~\cite{he_NC2020}. 
In presence of strain, the primitive lattice vectors and hence the reciprocal lattice vectors get distorted. For a given strain matrix ${\mathcal E}$, (which satisfies ${\mathcal E}^T={\mathcal E}$ with $T$ denoting the transpose), the real space vectors distort as ${\bm r} \to (1 + {\mathcal E}) {\bm r}$ and the reciprocal vectors as ${\bm k} \to (1 - {\mathcal E}^T) {\bm k}$. We obtain the strained \moire lattice vectors as ${\bm G}_{m}^i = {\bm \beta}_i^{(1)} - {\bm \beta}_i^{(2)}$ where ${\bm \beta}$ {represents the modified reciprocal lattice vectors due to combined effect of rotation and strain.} 
More details of the impact of strain on the electronic properties of TDBG are summarized in Appendix~\ref{BS_strain}.

To calculate the strained band structure we consider the uni-axial strain of strength ${\mathcal E}$ at an angle $\phi$ relative to zigzag direction as~\cite{bi_PRB2019_designing, pantaleon_PRB2021_tunable, sonia_PRBL2021_twist}
\be \label{strain}
{\mathcal E} = \varepsilon 
\begin{pmatrix}
-\cos^2 \phi + \nu \sin^2 \phi & (1+\nu ) \sin \phi \cos \phi \\
(1+\nu) \sin \phi \cos \phi & -\sin^2 \phi + \nu \cos^2 \phi
\end{pmatrix}~.
\ee
Equation~\eqref{strain} represents the system when it is more stretched in one direction and less stretched in the perpendicular direction.
With this strain matrix, the calculated electronic structure of TDBG for $\varepsilon=0.2\%$ strain (with $\phi=0$) is shown in Fig.~\ref{fig2}(b). We find that the strain can lift the degeneracy between the flat bands observed in the  absence of a perpendicular electric field ($\Delta=0$). 

More importantly, even a small but finite strain significantly reduces the symmetry of the system. 
In Fig.~\ref{fig2}(c) and (d), we highlight the breaking of the $C_3$ rotational symmetry, one of the key symmetry of the hexagonal lattice structure, due to strain. We convey this by plotting the energy dispersion along three different $C_3$-symmetric paths. We show that in absence of strain [see Fig.~\ref{fig2}(c)] the energy dispersion in the three marked paths lie on top of each other. In contrast, the presence of strain in the Fig.~\ref{fig2}(d) lifts this degeneracy, clearly indicating the breakdown of the $C_3$ symmetry. 

\begin{figure*}[t!]
    \centering
    \includegraphics[width=2.\columnwidth]{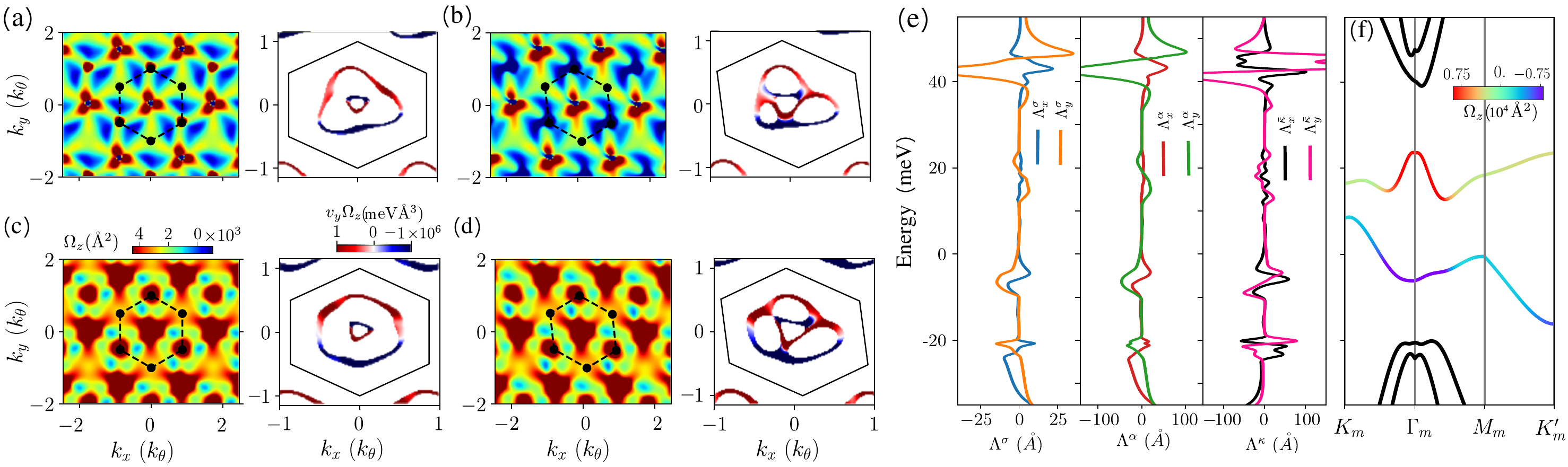}
    \caption{(a)-(d) The distribution of the Berry curvature ($\Omega_z$) and the distribution of the integrand of the Berry curvature dipole ($\Omega_z v_y$) in the Brillouin zone at a given energy (Fermi energy). (a) and (c) are in absence of strain but for electric field values of $\Delta=25$ meV and $\Delta=37$ meV, respectively. (b) and (d) are plotted for the same electric field strength but with $0.2\%$ strain. (e) The variation of the Berry curvature dipole, the thermoelectric counterpart of the Berry curvature dipole and the thermal counterpart of the Berry curvature dipole with energy at $\Delta=11$ meV. (f) The Berry curvature weighted band dispersion at $\Delta=11$ meV. Clearly the Berry curvature dipole and its counterparts become significant near the hotspots of the Berry curvature in the band dispersion.}
    \label{fig3}
\end{figure*}

\section{Nonlinear anomalous Hall effects in TDBG}
\label{NLAHE_TDBG}

Having explored the electronic band structure of TDBG, we now focus on the electric, thermoelectric and thermal NLA Hall effects~\cite{du_NRP2021_nonlinear, ortix_AQT2021_nonlinear}. It has been recently discovered~\cite{sodemann_PRL2015_quantum} that SIS broken systems possess a second-order nonlinear Hall charge current, 
which is finite even in the presence of TRS. This has been termed as the NLA Hall effect. The corresponding conductivity $\sigma_{abc}$ relating the NL current to the in-plane electric field, ${\bm E}$ ($j^{(2)}_a = \sigma_{abc} E_b E_c$, with $a$, $b$ and $c$ as spatial indices) is given by
\be \label{sigma_abc}
\sigma_{abc} = \epsilon_{abd} \dfrac{e^3 \tau}{\hbar^2} \Lambda_{dc}^\sigma~.
\ee
Here, $-e$ is the electronic charge, $\epsilon_{abd}$ is the anti-symmetric Levi-civita tensor, $\tau$ is the scattering time and $\Lambda_{dc}^\sigma$ is the BCD which is given by, 
\be \label{Lambda}
\Lambda^\sigma_{dc} = - \sum_n \int [d{\bm k}]\dfrac{\partial f_0^n}{\partial k_c} \Omega^n_d~.
\ee
In Eq.~\eqref{Lambda}, $f^n_0$ is the Fermi distribution function of the $n$-th band and $[d{\bm k}]$ stands for $d{\bm k}/(2 \pi)^D$, where $D$ represents the spatial dimension of the system and $\Omega^n_d$ is the $d$-th component of Berry curvature vector. It is defined as $\Omega^n_d= \frac{1}{2} \epsilon_{dac} \Omega_{ac}^n$, where
\be 
\Omega^n_{ac} = - 2 {\rm Im} \sum_{m \neq n  } \dfrac{ \langle  u_n |\partial_{k_a} {\mathcal H} | u_m \rangle \langle  u_m |\partial_{k_c} {\mathcal H} | u_n\rangle}{(\epsilon_n - \epsilon_m)^2} ~.
\ee
Here, $u_n$ is the periodic part of the Bloch wave-function with ${\mathcal H} | u_n \rangle = \epsilon_n | u_n \rangle $. The discovery of electrical NLA Hall effect was followed  up by predictions where a longitudinal temperature gradient is shown to generate second-order NLA Hall charge current as well as NLA Hall heat current. The phenomena of temperature gradient induced NLA Hall charge current, $j_a^{(2)}=\alpha_{abc}(\nabla_b T \nabla_c T)$, has been named as NLA Nernst effect~\cite{yu_PRBR2019_topological, zeng_PRB2019_nonlinear}. The NLA Nernst coefficient can be expressed as $\alpha_{abc} = \epsilon_{abd} \frac{e \tau k_B^2}{\hbar^2} \Lambda_{dc}^\alpha$ with $\Lambda_{dc}^\alpha$ being the thermoelectric counterpart of the BCD. Similarly, the NLA heat current generation due to temperature gradient $j_{Q,a}^{(2)}=\bar \kappa_{abc}(\nabla_b T \nabla_c T)$ is called the NLA Righi-Leduc effect~\cite{zeng_PRR2020_fundamental}. The corresponding NL conductivity can be expressed in terms of the thermal counterpart of the BCD $\Lambda_{dc}^{\bar \kappa}$ as $\bar \kappa_{abc} = \epsilon_{abd} \frac{\tau k_B^3 T} {\hbar^2} \Lambda_{dc}^{\bar \kappa}$. The expressions of all the NLA Hall conductivities and the related generalization of the Berry curvature dipoles are  summarized in Table.~\ref{table_1}.
\begin{table} 
\begin{tabular}{c c }
\hline
\hline
\rule{0pt}{3ex}
NL conductivity~~~  & Berry curvature dipoles  \\ 
\hline
\rule{0pt}{4ex}   
 $\sigma_{abc} = \epsilon_{abd} \dfrac{e^3 \tau}{\hbar^2} \Lambda_{dc}^\sigma$~~~ & $\Lambda^\sigma_{dc} = - \sum_n \int [d{\bm k}]\dfrac{\partial f^n_0}{\partial k_c} \Omega_d^n$
\\ [2ex]
 $\alpha_{abc} = \epsilon_{abd} \dfrac{e \tau k_B^2}{\hbar^2} \Lambda_{dc}^\alpha$~ & $\Lambda^\alpha_{dc}=-\sum_n\int [d{\bm k}] \dfrac{(\epsilon_n - \mu)^2}{(k_B T)^2} \dfrac{\partial f^n_0}{\partial k_c} \Omega^n_d$
\\ [2ex]
 $\bar \kappa_{abc} = \epsilon_{abd} \dfrac{\tau k_B^3 T}{\hbar^2} \Lambda_{dc}^{\bar \kappa}~ $ & $~~\Lambda^{\bar \kappa}_{dc} = -\sum_n \int [d{\bm k}] \dfrac{(\epsilon_n - \mu)^3}{(k_B T)^3} \dfrac{\partial f_0^n}{\partial k_c} \Omega_d^n $
 \\ [2ex]
\hline
\hline
\end{tabular}
\caption{The family of the nonlinear Hall conductivity (electric, thermoelectric and thermal) and the corresponding generalizations of the Berry curvature dipoles. The NLA Hall effect ($\sigma_{abc}$) is quantified by the BCD $\Lambda_{dc}^\sigma$, the NLA Nernst effect is quantified by $\Lambda_{dc}^\alpha$ and the NLA thermal Hall effect is quantified by $\Lambda_{dc}^{\bar \kappa}$. Clearly, all three of these are Fermi surface effects.}
\label{table_1}
\end{table}


Similar to other second-order NL transport phenomena, the NLA Hall effects vanish in systems with the SIS. This can be seen from the expressions of Table.~\ref{table_1} using the fact that in presence of SIS, both the energy and Berry curvature are even functions of the crystal momentum. Interestingly, the breaking of SIS is not sufficient to generate the NL Hall responses and further reduction in the crystal symmetries is  required \cite{sodemann_PRL2015_quantum}. 
For a detailed crystal symmetry analysis for materials which can host a finite BCD, we refer the readers to Ref.~[\onlinecite{sodemann_PRL2015_quantum}].
Furthermore, from the expressions of the NLA Hall effects given in Table~\ref{table_1}, the different NLA conductivity coefficients are expected to be related on the basis of the Sommerfeld expansion. In the low temperature limit, $\mu \gg k_B T$, the thermoelectric counterpart (Nernst) satisfies the relation $\alpha_{abc} (\mu) \propto \sigma_{abc} (\mu)$ while the NL thermal Hall coefficient satisfies ${\bar \kappa}_{abc}(\mu)= \frac{\partial}{\partial \mu} \sigma_{abc}(\mu)$.

In TDBG, the TRS is present and the SIS is broken. Broken SIS combined with the absence of even-fold rotational symmetries, guarantees the presence of a finite Berry curvature~\cite{wu_PRL2018_theory}. Since TDBG is a 2D-system, the Berry curvature has only one component perpendicular to the 2D plane (chosen to be the $z$-component). The distribution of the Berry curvature in the $K$-valley, in absence and in presence of strain are shown in Fig.~\ref{fig3}(a)-(d) for two values of the perpendicular electric field. We emphasize  that the presence of TRS in TDBG guarantees that $\Omega_{z, K}({\bm k}) = - \Omega_{z, { K}'}(-{\bm k})$. The non-zero Berry curvature combines with the velocity distribution over the BZ to dictate the NL responses at the Fermi surface. Interestingly, we find that although each valley possesses  large Berry curvature, the BCD and its thermoelectric and thermal counterparts are identically zero in each valley of pristine TDBG due to the presence of $C_{3}$ symmetry. However, in  any  realistic system, strain breaks the $C_3$ symmetry to generate a finite BCD and its thermoelectric and thermal counterparts. 
The $K$-valley distribution of the Berry curvature multiplied with velocity at the first conduction band has been highlighted in Fig.~\ref{fig3}(a)-(d) both in absence and presence of $C_3$ symmetry. The asymmetry of Fig.~\ref{fig3}(b) and (d) promotes a net dipole within the \moire unit cell. Note that the presence of TRS implies ${\bm v}_{n, K} ({\bm k})= {\bm v}_{n ,{ K}'}(-{\bm k})$. Consequently, the Berry curvature dipoles of the two valleys are identical.

In 2D systems like the TDBG, where the Berry curvature has only one component, the Berry curvature dipoles $\Lambda^{\sigma/\alpha/{\bar \kappa}}_{dc}$ are pseudo-tensors. Consequently it can be represented by two components of a vector decided by the direction of the velocity (subscript $c$). In Fig.~\ref{fig3}(e) we have plotted the Berry curvature dipoles $\Lambda^\sigma_{x/y}$, $\Lambda^\alpha_{x/y}$ and $\Lambda^{\bar \kappa}_{x/y}$ of TDBG as a function of energy in presence of $0.2\%$ strain for perpendicular electric field $\Delta=11$ meV. This is supplemented by the energy dispersion weighted by the Berry curvature along some high-symmetry points in Fig.~\ref{fig3}(f). 
In the BCD distributions along the energy axis, broadly four regions of non-zero BCD and its counterparts can be seen. Each of these regions are separated by the band gaps. Among them two regions (in the middle) can be attributed to the flat conduction and valence bands, while the other two (on top and bottom) originate from the higher moir\'{e} bands. 

From the plots we find that the BCD in the flat bands is of the order of magnitude $\sim 10$\AA. The thermoelectric and thermal counterparts of BCD have larger magnitudes $\sim 50$\AA  ~and $\sim 75$\AA, respectively. As expected, the thermoelectric counterpart of the BCD follows the same trend as the BCD. However, the thermal counterpart provides more subtle feature that are dictated by the derivative with respect to energy. We note that, although the BCD and its counterparts as a whole concentrate near certain regions, the $x$-component and the $y$-component of the Berry curvature dipoles have different features. 

\begin{figure}[t!]
    \centering
    \includegraphics[width=\columnwidth]{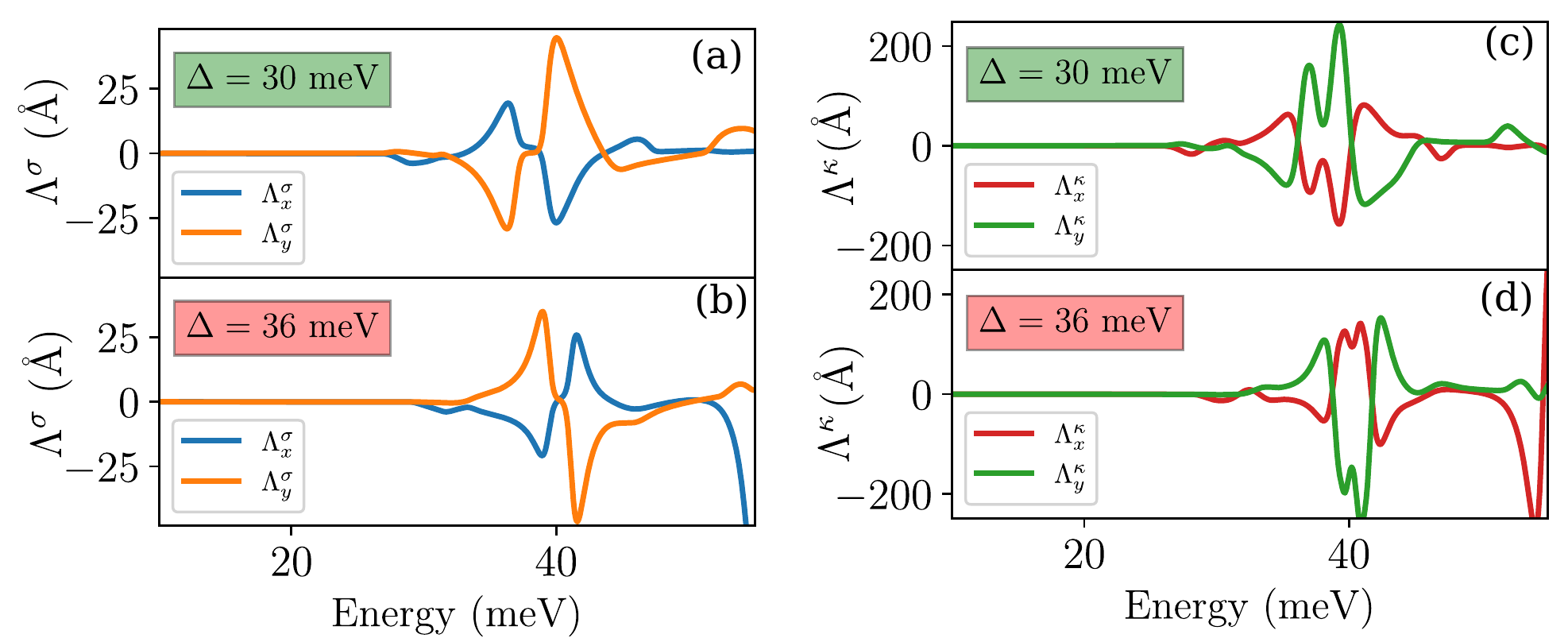}
    \caption{(a)-(b) Perpendicular electric field induced sign change in the Berry curvature dipole, $\Lambda^{\sigma}_{x,y}$. For $\Delta=30$ meV the $\Lambda^{\sigma}_{y}$ ($\Lambda^{\sigma}_{x}$) has a dip (peak) first and then a peak (dip) with increasing energy. This trend gets reversed for $\Delta=36$ meV. (c)-(d) Highlight a similar sign reversal in the Berry curvature dipole counterpart for the thermal  conductivity $\Lambda^{\bar \kappa}_{x,y}$.}
    \label{transition}
\end{figure}

To highlight impact of the perpendicular electric field ($\Delta$) on the nature of the BCD and its thermal counterparts, we have plotted them for two different values of $\Delta$ in Fig.~\ref{transition}. Specifically, we have plotted $x$- and $y$-components of $\Lambda^\sigma$ and $\Lambda^{\bar \kappa}$ on the conduction band side for $\Delta=30$ meV and $\Delta=36$ meV. Interestingly, we find that the distribution of the BCD and its thermal counterpart changes sign for a fixed energy (or for a fixed filling factor) when we go from $\Delta=30$ meV to $\Delta=36$ meV. As we will show in the next section, this sign change in the nonlinear conductivity is extremely important in systems with TRS and it captures phase transition in the valley-topology (change in the valley-Chern number) of such systems.

\section{Signature of topological phase transition  in nonlinear anomalous Hall responses}
\label{tplgy_TDBG}

In this section, we connect the NLA Hall responses with the topological phase transition in the electronic band structure of TDBG.
It has been shown~\cite{koshino_PRB2019_band, chebrolu_PRB2019_flat} that the TDBG possesses valley-Chern number, which is the valley analog of the Chern number usually defined in Chern insulators (TRS broken systems). While the Chern number is calculated by integrating the Berry curvature over the whole Brillouin zone, the valley-Chern number is calculated by integrating the Berry curvature near the valleys. Specifically, we define the valley-Chern number for the  $n$-th band as 
\be
{\cal C}_n^{\rm V} = \frac{1}{2 \pi} \int_{\rm valley} d{\bm k} \Omega_{ z}^n ~.
\ee
Strictly speaking, the concept of valley-Chern number is an approximate idea that depends only on the integration near a specific valley and does not map on to the full BZ~\cite{zhang_PNAS2013_valley}. However, since the continuum model of the moir\'e systems provides valley specific BZ and electronic structure, we can calculate the valley-Chern number in moir\'e systems just like the usual Chern number. We add here that in  presence of TRS in TDBG, the total Chern number obtained by adding the contribution of both the valleys is always zero.

\begin{figure*}[t!]
    \centering
    \includegraphics[width=2.0\columnwidth]{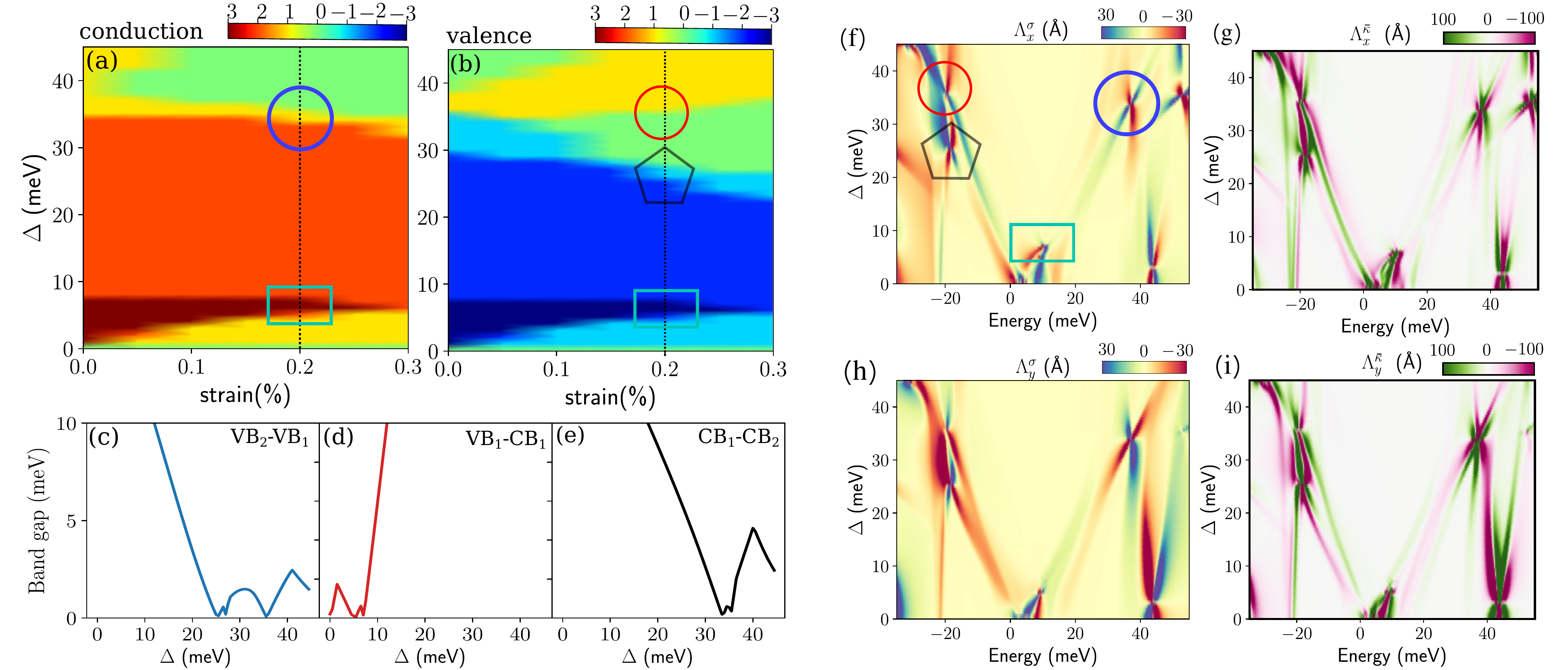}
    \caption{The valley-Chern phase diagram of the (a) first conduction band and (b) the first valence band as a function of  strain and the perpendicular electric field ($\Delta$). For the valence band, the valley Chern number varies from $-3$ to $1$ while for conduction band it varies from $0$ to $3$. The minimum direct band gap between the second valence band (VB) and the first  valence band, first VB and the first conduction band (CB), and first CB and second CB are shown in (c), (d) and (e) respectively. Clearly, each of the topological phase transition is associated  with the bands touching each other at a specific ${\bm k}$ point. (f)-(i) The variation of the $x$- and $y$- components of the Berry curvature dipole and its thermal counterparts with energy and $\Delta$. Panels (f) and (h) show the Berry curvature dipoles which has a butterfly-like structure with four lobes in vicinity of the topological phase transition point. Panels (g) and (i) present the thermal counterpart of the BCD which has a butterfly-like structure with eight lobes, at the location of the topological phase transition.}
    \label{fig5}
\end{figure*}
Interestingly, the valley-Chern number in TDBG can be changed  by applying a perpendicular electric field inducing a  topological phase-transition of valley-Chern type~\cite{ezawa_PRBR2013_topological,zhang_PNAS2013_valley}. This is different from the familiar topological phase-transitions in Chern insulators~\cite{haldane_PRL1988_model, ren_RPP2016_topological} and defined in systems with TRS where the Chern number vanishes. In Fig.~\ref{fig5}(a)-(b), we present the variation of the valley-Chern number of the first conduction and first valence bands in the $K$-valley on changing the perpendicular electric field ($\Delta$) and strain ($\varepsilon$). We find  multiple phase transitions in TDBG on varying strain and the perpendicular electric field. This  is consistent with earlier reports of the valley-Chern phase transitions in TDBG with perpendicular electric field in Refs.~[\onlinecite{wang_PRB2021_phase, mohan_PRB2021_tri}], in absence of strain. Here, we find that the strain impacts the phase-transition boundaries significantly and also induces new phases for higher values of the electric field strength [see green region in Fig.~\ref{fig5}(b)]. We note that the valley-Chern phase $(2,-2)$, representing the conduction and valence band valley-Chern numbers respectively, is the dominant phase in TDBG. In addition, there are many other phases with valley-Chern number ranging from $-3$ to $3$ depending on the specific choice of the parameters. 

Like the usual phase-transition in Chern insulators, the valley-Chern number changes in TDBG are also associated with band gap closing at a specific ${\bm k}$ point. To show this explicitly we have explored the evolution of the band movement in Fig.~\ref{fig5}(c)-(e). Specifically, we have plotted the minimum of direct band-gaps as a function of $\Delta$ for the four lowest energy bands near the charge neutrality point (two flat bands along with the two \moire bands enveloping them), for strain $\varepsilon=~0.2\%$. It is evident from Fig.~\ref{fig5}(d) that the valley-phase transition for small perpendicular electric field ($\Delta \approx 8$~meV) in both the bands are associated with band gap closing between the pair of flat bands. Similarly, the phase-transitions of the conduction band for relatively higher value of perpendicular electric field ($\Delta \approx 28$~meV) can be attributed to the band gap closing between the first and second conduction band as highlighted in Fig.~\ref{fig5}(e). The change in the valley Chern number of the valence flat band, for larger field strength, are associated with successive closing of band gaps between first and second valence bands, see Fig.~\ref{fig5}(c). 

A peculiarity of these topological phase transitions is that they are difficult to detect experimentally. This is  because there is no net circulating edge state as the total Chern number, summed over both the two valleys, is zero. We show below that the NLA Hall measurements offer a unique probe to detect these valley-Chern phase transitions. At the heart of this idea is the fact that band geometric quantities dictate the band topology, and in this case of time reversal symmetric system they also manifest in the NLA Hall transport phenomena. This can be easily seen via the simple model calculation for gapped graphene presented in Appendix~\ref{tilted_massive} and summarized in  Fig.~\ref{fig_SM}. 

To highlight the connection between the valley-Chern phase transition and the NLA Hall responses in TDBG, we show the variation of the BCD ($\Lambda^\sigma_i$) with energy and perpendicular electric field ($\Delta)$ space for $\varepsilon= 0.2$\% in Figs.~\ref{fig5}(f) and ~\ref{fig5}(h). The corresponding thermal counterpart is shown in Figs.~\ref{fig5}(g) and ~\ref{fig5}(i). The NL thermoelectric counterpart of the BCD is simply proportional to the BCD, and not presented separately. The key features in the $\Lambda^\sigma_i$ plot are the appearance of butterfly-like patterns with four lobes. In the $\Lambda^{\bar \kappa}_i$ plot the corresponding patterns are replaced by butterfly-like patterns with eight lobes. 
The doubling of the number of lobes in $\Lambda^{\bar \kappa}_i$ is a consequence of the fact that $\Lambda^{\bar \kappa}_i$ is the energy derivative of the $\Lambda^\sigma_i$~[see Fig.~\ref{fig1} (c)].

A careful analysis reveals that the butterfly-like structures accompany  the topological phase transition, and these are a unique signature of the transition. 
To highlight this, we have encircled two regions in the topological phase diagram of the conduction band [see Fig.~\ref{fig5}(a)], one near $\Delta=8$ meV and the other near $\Delta=34$ meV. Corresponding to these two points we have encircled two regions in the BCD phase-space plot [see Fig.~\ref{fig5}(f)], which show butterfly-like structures. Similarly, we have encircled three phase-transition points in the valence band valley-Chern phase diagram in vicinity of $\Delta=8$ meV, $\Delta=28$ meV and $\Delta=34$ meV. The corresponding regions having the butterfly-like structures are also marked in the BCD phase-space diagram of Fig.~\ref{fig5}(f). 
Experimentally, we can fix the filling factor in vicinity of a Berry curvature hotspot (near a band edge), and measure the variation of the BCD with change in the perpendicular electric field. The butterfly structure described above, will feature in such experiments via the sign reversal in the NLA Hall conductivities across the phase transition, as shown in Fig.~\ref{transition}. This establishes that the sign reversal in the NLA Hall conductivities can be used as an experimental probe for the topological phase transition. The physical origin of the butterfly-like structure near the phase transition is a bit subtle, and it can be understood as follows. If we imagine a horizontal line through the center of the butterfly, then each of the lower two wings (lobes) of the butterfly corresponds to the BCD contribution of the two bands which touch each other at the phase-transition  point. The corresponding colors represent opposite BCD before the transition. On the other hand, the upper two lobes of the butterfly-like structure are the BCD contribution of the same bands after the phase transition and the change in color at the same energy (imagine a vertical line) represents the phase transition. For instance, consider the butterfly around $\Delta=$ 34~meV and $\mu \approx$ 35~meV. The lower left (right) lobe is due to the first (second) conduction bands before phase-transition. The upper left (right) lobe represents the same band BCD contributions, but after the phase-transition.

From the above discussion, we expect a butterfly-like structures in the BCD diagram for each of the phase-transition points. However, there are few more subtleties worth mentioning  here. The BCD and its thermal counterparts are Fermi surface phenomena and hence the total contribution at a given doping (or chemical potential) originates from all the bands that cut the Fermi surface. This implies  that the contribution from the individual bands to the NLA Hall effects, which we connect to the valley-Chern number of the corresponding band, will be uniquely  distinguishable only for isolated band structures. However, in real systems multiple other energy bands may cross the Fermi energy, jeopardizing the appearance of the perfect BCD butterflies in the energy-doping plane. Furthermore, the net contribution from all the contributing bands can sometimes be small, making it difficult to distinguish the perfect butterfly structure. For instances, at $\Delta \sim 8$ meV we do not observe a complete butterfly for the conduction band phase-transition.

\section{Experimental implications}
In the last section, we have shown that the topological phase-transitions can be probed through the sign change in different BCD induced NLA Hall coefficients. However, in an experimental set up, the total NLA Hall response can have additional contributions from skew-scattering and side jump processes along with the BCD contribution~\cite{Du2019D}. To separate out the BCD induced contribution of the NLA Hall signal from the rest and observe the sign change due to topological phase-transition is challenging and require some experimental sophistication. This can be done as follows. The different contributions of the NLA Hall signal scale differently with the scattering timescale or with the linear DC conductivity. By studying the scaling of total NLA Hall response with the Drude conductivity, the BCD dependent contributions can be extracted from the total NLA Hall response. In particular, the perpendicular electric field provides a useful knob to vary the conductivity at a fixed temperature to study the scaling in TDBG and detect topological transition from the sign-reversal of BCD~\cite{sinha_NC2022_berry}. A fixed temperature ensures that the scaling parameters are not tuned with temperature~\cite{Xiao2019S} and helps in accurate estimation of BCD. We note that measuring the other two counterparts of BCD (thermoelectric and thermal) requires control over heat current generation and detection and this is relatively harder to measure experimentally. However, recent advancements in thermal measurements~\cite{Srivastav2019U, Dhara2011T} may provide a way forward in the near future.


\section{Conclusion}
\label{conclusion}
To summarize, we have investigated the NLA Hall effect, the NLA Nernst effect, and the NLA thermal Hall effect in TDBG. We highlight that the breaking of the $C_3$ rotational symmetry in TDBG, by strain, is crucial for having finite NL Hall responses in \moire systems with trigonal symmetry. We find that the NLA Hall conductivities for charge and heat transport are maximum near the band edges, which are also the hotspots of the Berry curvature. 
Additionally, we show that the tunability of the band structure of TDBG through the perpendicular electric field can be exploited to tune the NLA Hall responses. 
Further, we demonstrate that the measurement of NLA Hall effects can be used as an experimental tool to probe  topological phase-transitions of the valley-Chern type in time reversal symmetric systems. Specifically, we find that each topological transition is accompanied with a sign change in the NLA Hall coefficients, as reflected by the butterfly-like patterns in our calculations. Our study establishes the NLA  Hall responses as an effective probe for detecting topological phase-transitions in \moire superlattices and in other systems.

\section{Acknowledgment}
\noindent A.C acknowledges Indian Institute of Technology, Kanpur and  Science and Engineering Research Board (SERB) National Postdoctoral Fellowship (PDF/2021/000346), India for financial support. We acknowledge the Science and Engineering Research Board [for projects MTR/2019/001520, and CRG/2018/002400] and the Department of Science and Technology (DST, for project DST/NM/TUE/QM-6/2019(G)-IIT Kanpur) of the Government of India for financial support. M.M.D. acknowledges DST SUPRA SPR/2019/001247 grant along with the Department of Atomic Energy of Government of India 12-R$\&$D-TFR-5.10-0100 for support. We thank CC-IITK for providing the high performance computing facility.

\appendix
\section{Band structure of TDBG in absence of strain} 
\label{BS_wo_strain}

In this appendix, we construct the Hamiltonian of TDBG. We start with an initial brief review of the Bernal stacked bilayer graphene Hamiltonian, which is the building block of the TDBG. The Brillouin zone (BZ) of the AB-stacked bilayer graphene is identical to that of a monolayer graphene. The primitive lattice vectors ${\bm a}_1 = a(1,0)$ and ${\bm a}_2 = a(1/2, \sqrt{3}/2)$ yield the reciprocal lattice vectors to be ${\bm b}_1^{(0)} = \frac{4 \pi}{\sqrt{3} a} (\sqrt{3}/2 , - 1/2)$ and ${\bm b}_2^{(0)} =\frac{4 \pi}{\sqrt{3} a} (0, 1)$. Here, $a$ is the lattice constant which is $\sqrt{3}$ times of the carbon-carbon bond length $d=1.42$ \AA. The coordinates of the vertices of the hexagonal first BZ are ${\bm K}_{\xi}^{(0)} = \xi (2 {\bm b}_1^{(0)} + {\bm b}_2^{(0)})/3$ with $\xi$ being the valley index. Near these BZ corners the electronic band structure of bilayer graphene comprises of massive Dirac dispersion~\cite{edward_RPP2013_the, rozhkov_PR2016_elec} and it can be described by an effective four band low energy model. Including the effects of hexagonal warping, the Hamiltonian near the $K$-valley can be expressed in terms of the fermion operators of the $A$ and the $B$ sublattice of the top and the bottom layers, $[\{c^t_A(k), c^t_B(k)\},\{ c^b_A(k), c^b_B(k)\}]$, as
\be \label{ham_blg}
H(k)=\begin{pmatrix}
h_k^t & t_k \\ t_k^\dagger & h_k^b
\end{pmatrix}.
\ee
Here, the block diagonal matrices $h_k^{t/b}$ represents the massive Dirac Hamiltonian of the top and bottom monolayers and $t_k$ represents the effect of inter-layer hopping. The corresponding matrices are
\be
h_k^{(t/b)} = 
\hbar v_0 {\bm \sigma} \cdot {\bm k}  + \frac{\delta}{2} ( \mathbb{I} \mp \sigma_z),~~ t_k= \begin{pmatrix}
-\hbar v_4 \pi^\dagger & -\hbar v_3 \pi \\ \gamma_1 & -\hbar v_4 \pi^\dagger
\end{pmatrix}
\ee
with $\pi\equiv k_x + i k_y$. In the Hamiltonian, different intra-layer and inter-layer couplings have been introduced through the hopping parameter $\gamma_i$ or equivalently by $v_i = \sqrt{3} |\gamma_i| a /(2 \hbar)$. The nearest neighbor  intra-layer coupling between the $A$ and the $B$ sublattice is represented by the parameter $v_0$. The inter-layer intra-dimer coupling is represented through $\gamma_1$. Parameters $\gamma_3$ and $\gamma_4$ are the couplings between the inter-layer non-dimer sites coupling and the inter-layer coupling between dimer and non-dimer sites respectively~\cite{mohan_PRB2021_tri}.
For our calculations we consider $\delta=15 $ meV, $\gamma_0 = -3.1$ eV, $\gamma_3=283$ meV and $\gamma_4=138$ meV. 

A schematic of a moir\'e pattern is shown in Fig.~\ref{fig2}(a) where the right-side arrow orientations (red, blue and green) represent the ${K}$-valley, while the left-side represents the $K'$-valley. 
The reciprocal lattice vectors of the \moire lattice are obtained as ${\bm G}_{m}^i={\bm b}_i^{(1)}-{\bm b}_i^{(2)}$, with the rotated reciprocal lattice vectors of each bilayer being specified by ${\bm b}_1^{(l)} = {\cal R}(\mp \theta/2) {\bm b}_1^{(0)}$ with $\mp$ for bilayer $l=1,2$, respectively. Using this we obtain the pair of primitive \moire lattice vectors to be ${\bm G}_{m}^1=\frac{8 \pi}{\sqrt{3}a} \sin \frac{\theta}{2}(-1/2, \sqrt{3}/2)$ and ${\bm G}_{m}^2=\frac{8 \pi}{\sqrt{3}a} \sin \frac{\theta}{2}(1/2, \sqrt{3}/2)$. Using the low energy Hamiltonian [Eq.~\eqref{ham_blg}] for each lattice points, vertices of the small hexagons in Fig.~\ref{fig2}(a), and the moir\'e hopping matrix, we construct the continuum Hamiltonian. A certain cut-off in the reciprocal space is used to truncate the lattice. 
The smallest TDBG Hamiltonian for the $K$-valley can be written as
\be 
{\mathcal H} = 
\begin{pmatrix}
	h_{k,t}^++\Delta_t^+ & t_k^{+} & 0 & 0 \\
	{t^+_k}^\dagger & h_{k,b}^{+}+\Delta_b^+ & T & 0 \\
	0 &  T^\dagger & h_{k,t}^-+\Delta_t^- &  t_k^-\\
	0 & 0 & {t_k^-}^\dagger  & h_{k,b}^{-}+\Delta_b^- \\
\end{pmatrix} .
\label{tdbg_ham}
\ee
Here, the superscripts on $h^{\pm}_{k,t/b}$ represents rotated Dirac Hamiltonian as $h^{\pm}={\mathcal R}(\mp \theta/2) {\bm k}\cdot{\bm \sigma}$ and $\Delta_{t/b}^{\pm}$ represents the effect of a perpendicular electric field. In Eq.~\eqref{tdbg_ham}, $T({\bm r})$ represents the moir\'e coupling matrix, which  connects the bottom layer of bilayer-$1$ to the top layer of bilayer-$2$. In this smallest TDBG Hamiltonian, only the nearest neighbor coupling  will be considered which are connected by the vectors ${\bm q}_b= \frac{8\pi}{3a} \sin \frac{\theta}{2}(0, -1)$ and  ${\bm q}_{tl}  =\frac{8\pi}{3 a} \sin \frac{\theta}{2}(-\sqrt{3}/2, 1/2)$ and  ${\bm q}_{tr} =\frac{8\pi}{3 a} \sin \frac{\theta}{2}(\sqrt{3}/2, 1/2)$. The moir\'e hopping matrices are given by 
\be
T({\bm r}) =  \sum_{j=b, tr,tl}T_{{\bm q}_j} e^{-i{\bm q}_j \cdot {\bm r}}~,~~~{\rm where}
\ee
\be 
T_b = 
\begin{pmatrix}
\omega & \omega' \\
\omega' & \omega
\end{pmatrix}
~~~~~~~~T_{tr/tl} = 
\begin{pmatrix}
\omega & \omega' e^{\mp i 2\pi/3} \\
\omega' e^{\mp i 2\pi/3} & \omega
\end{pmatrix}~.
\ee
Here, $\omega$  and $\omega'$ denote the diagonal and the off-diagonal hopping strengths, respectively. We emphasize here that an unequal $\omega$ and $\omega'$, specifically $\omega' > \omega$, is known to be crucial to match the calculated low energy band structure with the  experimentally observed spectral gap~\cite{koshino_PRB2019_band, chebrolu_PRB2019_flat}. 
In this paper, we consider $\omega'=106$ meV and $\omega=79$ meV~\cite{koshino_PRB2019_band, mohan_PRB2021_tri}.

\section{Electronic band structure of TDBG in presence of strain}
\label{BS_strain}

In this section of Appendix we highlight the details of the strain implementation in TDBG.
In presence of strain, the Dirac Hamiltonian of Eq.~\eqref{tdbg_ham} modifies to
\be 
h_{k,l} =  \hbar v_0 {\mathcal R}(\mp \theta/2)~[(\mathbb{I} + {\mathcal E}^T)  ] ({\bm k} - {\bm D}_{\xi}) \cdot (\xi \sigma_x, \sigma_y) + \frac{\delta}{2} ( \mathbb{I} \mp \sigma_z)~.
\ee
Here, the strain matrix operate over the position of the twisted Dirac points given by
\be
{\bm D}_\xi = (\mathbb{I} - {\mathcal E}^{T}) {\bm K}^i_\xi - \xi {\bm A}~,
\ee
with ${\bm A}$ representing the gauge field that has the dimension of reciprocal lattice vector. The appearance of the gauge field can be attributed to the fact that the strain causes the inter-atomic distance in each layer to become different in different directions. This results in the difference of hopping parameters which displaces the Dirac point from its original position. The gauge potential ${\bm A}$ in terms of the elements of the strain matrix is given by
\be 
{\bm A} = \dfrac{\sqrt{3}}{{2a}} \beta ({\mathcal E}_{xx} - {\mathcal E}_{yy} , -2 {\mathcal E}_{xy})~.
\ee
Here, $\beta=1.57$ and ${\mathcal E}_{ij}$ are the elements of the strain matrix.

Strain also modifies the lattice vectors and consequently the hopping matrices and the hopping vectors. We calculate the strained moir\'e vectors starting from un-rotated and un-strained lattice vectors. Following Refs.~\cite{he_NC2020}, we obtain the lattice vectors using ${\bm G}_{m}^{1, {\rm st}}=R_{-\frac{\theta}{2}} (1-{\mathcal E}^T){\bm b}_1 - R_{\frac{\theta}{2}} {\bm b}_1$ and ${\bm G}_{m}^{2, {\rm st}}=R_{-\frac{\theta}{2}} (1-{\mathcal E}^T){\bm b}_2 - R_{\frac{\theta}{2}} {\bm b}_2$ which yields
\begin{widetext}
\begin{subequations}
\bea
{\bm G}_{m}^{1, \rm st} &=&\dfrac{k_\theta}{4} \Big( 2\sqrt{3} - 3{\mathcal E}_{xy} - \sqrt{3} {\mathcal E}_{yy} - (3{\mathcal E}_{xx} +\sqrt{3} {\mathcal E}_{xy})\cot \frac{\theta}{2},~ -6 + 3{\mathcal E}_{xx} + \sqrt{3} {\mathcal E}_{xy} -(3 {\mathcal E}_{xy} + \sqrt{3} {\mathcal E}_{yy})\cot \frac{\theta}{2} \Big),~~~
\\
{\bm G}_{m}^{2, \rm st}
&=&\dfrac{k_\theta}{4} \Big( 2\sqrt{3} + 3{\mathcal E}_{xy} - \sqrt{3} {\mathcal E}_{yy} + (3{\mathcal E}_{xx} -\sqrt{3} {\mathcal E}_{xy})\cot \frac{\theta}{2},~
 6 - 3{\mathcal E}_{xx} + \sqrt{3} {\mathcal E}_{xy} +(3 {\mathcal E}_{xy} - \sqrt{3} {\mathcal E}_{yy})\cot \frac{\theta}{2} \Big).~~~
\eea 
\end{subequations}
Now, assuming the $K$-valley on the right hand side of the graphene hexagon, one can define ${\bm q}_{\rm b}={\bm K}_{-\theta/2} - {\bm K}_{\theta/2}$. In presence of strain we calculate the modified hopping vector ${\bm q}_{\rm b}^{\rm st}$ to be~\cite{he_NC2020}
\bea 
{\bm q}_b^{\rm st}&=&{\bm K}_{-\theta/2}^{\rm st} - {\bm K}_{\theta/2} = R_{-\frac{\theta}{2}} (1-{\mathcal E}^T){\bm K} - R_{\frac{\theta}{2}} {\bm K}=\dfrac{k_\theta}{2}\left( -{\mathcal E}_{xy} - {\mathcal E}_{xx} \cot \frac{\theta}{2},~ -2 + {\mathcal E}_{xx} -{\mathcal E}_{xy} \cot \frac{\theta}{2}\right).~~
\eea
The other two hopping vectors can be obtained from ${\bm q}_{\rm tr}={\bm G}_{m}^{2} + {\bm q}_{\rm b}$ and ${\bm q}_{\rm tl}=-{\bm G}_{m}^{2} + {\bm q}_{\rm b}$. In presence of strain each of these vectors are replaced by their strained counterparts which yields
\begin{subequations}
\bea
{\bm q}_{tr}^{\rm st}
&=&\dfrac{k_\theta}{4} \Big( 2\sqrt{3} + {\mathcal E}_{xy} - \sqrt{3} {\mathcal E}_{yy} + ({\mathcal E}_{xx} -\sqrt{3} {\mathcal E}_{xy})\cot \frac{\theta}{2},~
 2 - {\mathcal E}_{xx} + \sqrt{3} {\mathcal E}_{xy} +({\mathcal E}_{xy} -\sqrt{3} {\mathcal E}_{yy})\cot \frac{\theta}{2} \Big),
\\
{\bm q}_{tl}^{\rm st}&=&\dfrac{k_\theta}{4} \Big( -2\sqrt{3} + {\mathcal E}_{xy} + \sqrt{3} {\mathcal E}_{yy} + ({\mathcal E}_{xx} +\sqrt{3} {\mathcal E}_{xy})\cot \frac{\theta}{2},~ 2 - {\mathcal E}_{xx} - \sqrt{3} {\mathcal E}_{xy} +({\mathcal E}_{xy} + \sqrt{3} {\mathcal E}_{yy})\cot \frac{\theta}{2} \Big).
\eea 
\end{subequations}
\end{widetext}
The strain also modifies the hopping matrices. For simplicity, we assume that the amplitude of the hopping matrices remain unchanged and the phase-factor only changes. We calculate the change in the phase factor $\phi$ as~\cite{he_NC2020}
\begin{subequations}
\bea 
\phi_{tr} &=& \frac{2 \pi}{3} \Big( 1 + \sqrt{3} {\mathcal E}_{xx} {\mathcal E}_{xy} + \sqrt{3} {\mathcal E}_{xy} {\mathcal E}_{yy} - {\mathcal E}_{xy}^2 - {\mathcal E}_{yy}^2 \Big),~~~~~~
\\
\phi_{tl} &=& \frac{2 \pi}{3} \Big( 1 - \sqrt{3} {\mathcal E}_{xx} {\mathcal E}_{xy} - \sqrt{3} {\mathcal E}_{xy} {\mathcal E}_{yy} - {\mathcal E}_{xy}^2 - {\mathcal E}_{yy}^2 \Big).~~~~~~
\eea
\end{subequations}

\section{Topological phase transition in tilted massive Dirac Hamiltonian} \label{tilted_massive}

\begin{figure}[t!]
    \centering
    \includegraphics[width=\columnwidth]{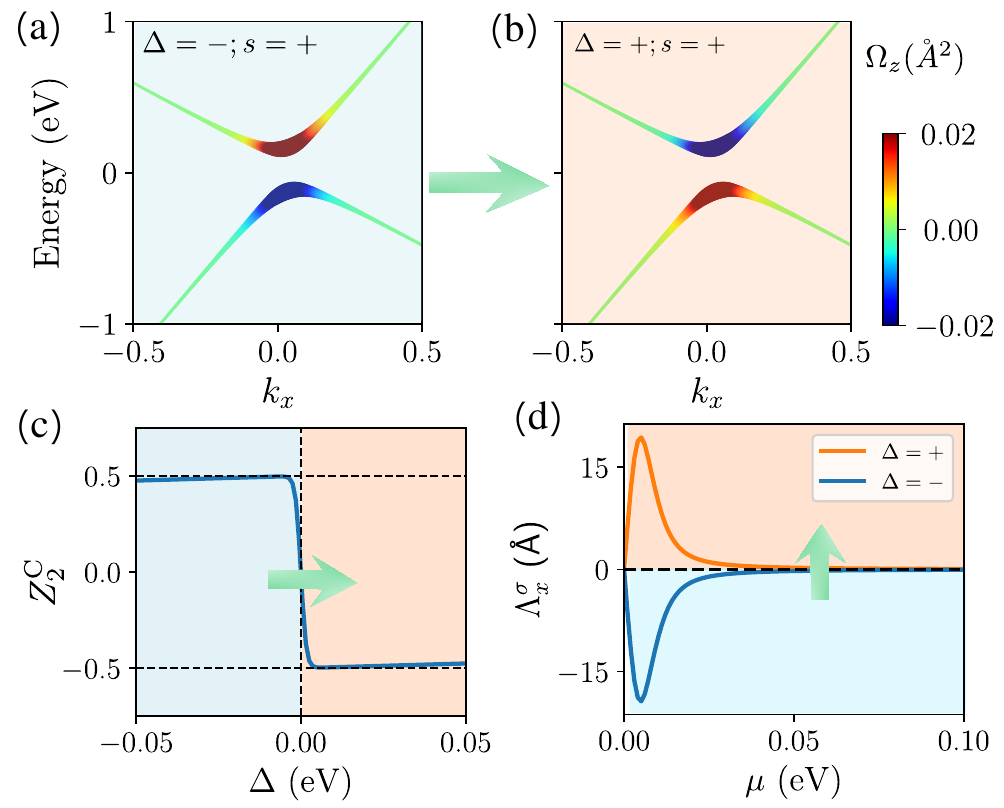}
    \caption{(a)-(b) The alteration of the Berry curvature distribution when the sign of the perpendicular electric field is reversed. (c) The sign change of the gap parameter is associated with a topological phase-transition of the valley-Chern type. (d) The topological phase transition can be experimentally detected by measuring the NLA Hall effect, whose sign changes across the phase transition. For the band structure and valley-Chern plot we have used $\hbar v_F=1.55$~eV/\AA, $v_t=0.4 v_F$ and $\Delta=0.3$ meV. The BCD is plotted at temperature $30$ K. }
    \label{fig_SM}
\end{figure}

In this Appendix, we convey the central idea of topological phase-transition of valley-Chern type employing a simplistic model for tilted massive Dirac Hamiltonian. For that we consider a model Hamiltonian as
\be \label{ham_p_1}
{\mathcal H}_s = \hbar v_F  (k_x \sigma_y - sk_y\sigma_x) + s \hbar v_t k_x + \Delta\sigma_z~.
\ee
Here, $v_F$ is the Fermi velocity and $s=\pm$ is the valley index and $\Delta$ is the gap parameter that captures the potential induced by the perpendicular electric field applied using the top and bottom gate and its sign can be reversed. The Berry curvature for this model is given by
\be \label{BC}
\Omega_z = \mp s \dfrac{\hbar^2 v_F^2\Delta}{2(\hbar^2 v_F^2k^2+\Delta^2)^{3/2}}~.
\ee
Here, the $\mp$ sign stands for the conduction and valence band.
Equation \eqref{BC} depends on the valley index $s$. For $s=+$ we have shown the Berry curvature weighted dispersion for $\Delta<0$ in Fig.~\ref{fig_SM}(a) and for $\Delta>0$ in Fig.~\ref{fig_SM}(b). A sign change in the Berry curvature distribution in the respective bands is evident.
Using the expression of Berry curvature, we can calculate the valley-Chern number. For that we need to first construct a definition of valley-Chern number for the massive Dirac low energy model (without properly defined BZ). The valley-Chern number may be defined in the infinite Dirac Fermi sea as
\be
{\mathcal C}_v = \dfrac{1}{2\pi} \int \Omega_z d{\bm k}~.
\ee
This expression can be computed numerically choosing the integration regime appropriately. However, for analytical calculation of the valley-Chern number we first calculate $\int \Omega_z f_0 d{\bm k} $ and then consider $\Delta \ll \mu$ the limit. Using this recipe for the isotropic Hamiltonian ($v_t=0$) we calculate the valley-Chern number for the conduction band with $\Delta>0$, to be
\be
{\mathcal C}_v = - \frac{s}{2} \left( 1 - \frac{\Delta}{\mu} \right)
\ee
Now in the infinite Fermi surface limit ($\Delta \ll \mu$) we get ${\mathcal C}_v =-s/2$. We note that the two valley have opposite opposite valley-Chern number. Since the valley-Chern numbers are opposite and sums up to zero, so a topological index can be defined as $Z_2=(C_+ -C_{-})/2$. Now, if we reverse the sign of the perpendicular electric field then the valley-wise Chern number gets altered. This electric field induced topological phase-transition has been highlighted in Fig.~\ref{fig_SM}(c). We emphasize that other than the perpendicular electric field, a topological transition can also occur by changing the stacking from AB to BA registry~\cite{zhang_PNAS2013_valley}. The consequence of the phase-transition on the NLA Hall effect has been highlighted in Fig.~\ref{fig_SM}(d).

\bibliography{NL_NE_moire.bib}

\end{document}